\documentclass[conference]{IEEEtran}
\IEEEoverridecommandlockouts
\usepackage{cite}
\usepackage{amsmath,amssymb,amsfonts}
\usepackage{algorithmic}
\usepackage{graphicx}
\usepackage{textcomp}
\usepackage{xcolor}
\def\BibTeX{{\rm B\kern-.05em{\sc i\kern-.025em b}\kern-.08em
    T\kern-.1667em\lower.7ex\hbox{E}\kern-.125emX}}

\usepackage[hidelinks]{hyperref}
\usepackage{randtext}
\usepackage{tikz}
\usetikzlibrary{shapes.geometric, arrows.meta, positioning, fit, backgrounds}
\usepackage{booktabs}
\usetikzlibrary{calc}
\usepackage{verbatim}  

\begin{document}

\title{From Code Generator to Engineering Companion: A Plant-aware Agent for PLC Programming}

\author{\IEEEauthorblockN{Joschka Kersting, Gesa Benndorf}
\IEEEauthorblockA{\textit{Fraunhofer IOSB-INA}\\
Lemgo, Germany \\
\randomize{\{joschka.kersting,~gesa.benndorf\}@iosb-ina.fraunhofer.de}}
}
\maketitle

\begin{abstract}
Programmable Logic Controller (PLC) coding remains a manual, IDE-centric task that demands constant context switching between documentation, project files, and code editors. Recent work has demonstrated that retrieval-augmented large language models can generate vendor-compliant Structured Text at high compilation rates \cite{pred}, yet these systems treat code generation as a one-shot pipeline: the engineer submits a query and receives code, with no persistent knowledge of the specific plant. We present a plant-aware conversational copilot that extends this paradigm in three directions: plant-specific context ingestion, a chat-first interaction model for explanation and debugging, and a graduated tool-access ladder from chat to deployment. The system is designed for real plant infrastructure and keeps the engineer in control of each step that produces real-world effects.
\end{abstract}


\footnote{\textcopyright~2026 IEEE. Personal use of this material is permitted.
Permission from IEEE must be obtained for all other uses, in any current
or future media, including reprinting/republishing this material for
    advertising or promotional purposes, creating new collective works, for
    resale or redistribution to servers or lists, or reuse of any copyrighted
    component of this work in other works.}

\section{Introduction}
Every production facility runs on control logic authored by automation engineers who navigate a demanding workflow: read the I/O specification, open the vendor IDE, locate the relevant function-block documentation, write Structured Text (ST) code, compile, fix errors, and repeat. Large language models have begun to reduce this friction. Prior work \cite{pred} demonstrated that a retrieval-augmented coding assistant for Mitsubishi Electric's GX Works3 environment can generate vendor-compliant ST code at high compilation rates using only a structured knowledge base of function-block definitions and documentation excerpts, without exposing proprietary code to cloud providers.

Yet that earlier system operates as a pipeline. The engineer submits a natural-language task, the model generates code, and the compilation server validates it. There is no persistent knowledge of the specific plant, let alone an iterative dialogue, and no path from chat to deployment.

The gap we address is not code generation per se, but the engineering interaction model. A copilot that knows the plant-specific I/O list, labels, and existing program fragments can generate code that is immediately appropriate for the target namespace. Moreover, in our setting we find that agents often only write correct code when they know the real-world setup of the plant. Together, these aspects form the difference between a generic coding tool and an engineering companion.
Unlike generic copilots (e.g., GitHub Copilot on a CODESYS MCP server), our contribution is not the chat mechanism itself but its coupling with plant-specific context retrieval and a domain-tailored, safety-graduated tool ladder.
\subsection*{Contributions}
\begin{itemize}
\item Plant-aware context ingestion that handles the specific plant's I/O list, device labels, module configuration, and existing program fragments.
\item A conversational interaction model for explaining, debugging, design discussion, and code modification.
\item Graduated tool access with four explicit levels: Chat, Code, Compile, \textbf{and} Deploy.
\end{itemize}

\begin{table*}[!hbt]

\caption{Comparison of related systems}
\centering

\begin{tabular}{|l|c|c|c|c|c|c|c|}
\hline
\textbf{System} & \textbf{Plant context} & \textbf{Code gen} & \textbf{Compile} & \textbf{Deploy} & \textbf{Chat-first} & \textbf{Multi-vendor} & \textbf{Target platform(s)} \\
\hline
LLM4PLC \cite{P1}            & -- & \checkmark & \checkmark &  -- & -- & -- & Siemens SCL \\
Agents4PLC \cite{P2}         & -- & \checkmark & \checkmark & -- & -- & \checkmark & IEC 61131-3 (generic) \\
AutoPLC \cite{P3}            & -- & \checkmark & \checkmark & -- & -- & \checkmark & Siemens SCL, CODESYS ST \\
Spec2Control \cite{S1}       & -- & \checkmark & --       & \checkmark & -- & -- & ABB \\
Siemens Copilot \cite{C1}    & -- & \checkmark & \checkmark  &  \checkmark & \checkmark & -- & Siemens TIA \\
Prior work \cite{pred}       & -- & \checkmark & \checkmark & -- & -- & -- & Mitsubishi iQ-R \\
\textbf{This work}           & \checkmark & \checkmark & \checkmark & \checkmark & \checkmark & -- & Mitsubishi iQ-R \\
\hline
\end{tabular}%

\label{tab:related}
\end{table*}

\section{Related Work}

\textbf{LLM-based PLC code generation.}  One line of work establishes iterative generation pipelines: a user-guided approach combines compiler, grammar-checker, and formal verification feedback, raising ST generation success on a manufacturing testbed from 47\% to 72\%~\cite{P1}.  A fully automated multi-agent successor adds RAG and Chain-of-Thought planning with dedicated syntax and verification agents; on a benchmark of 23 programming tasks (16 easy, 7 medium), the framework achieves a verifiable rate of 68\% on easy tasks and 42.3\% on medium tasks with GPT-4o~\cite{P2}.  A vendor-aware framework targets ST fragmentation across two proprietary platforms with a four-stage pipeline (planning, case retrieval, API recommendation, IDE-guided validation), reporting high compilation pass rates on a large benchmark at low average per-task cost~\cite{P3}.  A complementary fine-tuning direction applies iterative online Direct Preference Optimization, using compiler feedback and a specialist LLM as synthetic annotators to construct preference pairs without manual labelling~\cite{P4}; a separate system extends generation beyond ST to Sequential Function Charts, achieving 75--94\% syntactically valid output~\cite{P5}.

\textbf{LLM-based DCS engineering.}  A series of incremental contributions culminating in Spec2Control established a complete LLM workflow for Distributed Control System (DCS)  control logic: an initial study evaluated LLM capabilities across 100 PLC/DCS prompts in ten categories~\cite{koziolek2023chatgpt}; a follow-up integrated vendor function-block libraries via RAG~\cite{koziolek2024rag}; a third variant generated code from P\&ID diagrams using vision-capable models~\cite{koziolek2024img}.  Spec2Control presents an end-to-end LLM workflow to generate graphical IEC~61131-3 Function Block Diagrams directly from natural-language control narratives and deploy them to an automation controller~\cite{S1}.  It produces FBDs from proprietary ABB libraries and achieves 98.6\% correct strategy connections across 65 test cases from 10 narratives, with a 94--96\% reduction in engineering effort~\cite{S1}.  The key distinction from our work is the interaction model: Spec2Control is a batch workflow in which the engineer is consulted at low-confidence steps, whereas our system is conversational-first and leaves each tool action to explicit engineer initiation.

\textbf{Commercial engineering copilots.}  Several vendors have shipped conversational code assistance integrated into their proprietary IDEs~\cite{C1,C2,C3}, showing that major automation vendors are productizing this paradigm in commercial tools.  These products lack controlled evaluation datasets; Spec2Control characterises them as targeting textual notations with limited automation support~\cite{S1}.

\textbf{Human-in-the-loop and graduated autonomy.}  HITL patterns for LLM agents are well established in orchestration frameworks~\cite{T1,T2}, where agents pause before high-stakes tool calls pending explicit approval; human-guided development workflows following this pattern have been deployed in production software engineering~\cite{T3}.  Recent work formalises agent autonomy as a five-level spectrum of user roles---from full operator control to passive observation---providing a principled vocabulary for autonomy-level design~\cite{T4}.  Our contribution applies this design pattern to the industrial PLC domain, where an uncontrolled LLM-to-hardware action can cause physical damage and production downtime.

\textbf{Evaluation methodology.}  Existing work measures primarily compilation success and formal verification rates~\cite{P1,P2,P3}; Spec2Control adds GQM-structured quality and effort metrics~\cite{S1}.  For conversational copilots, however, metrics capturing interaction efficiency (turns to task completion), error distribution by correction cause, and cognitive load remain absent---gaps that our planned evaluation aims to address.

\section{System Design}
The system extends the RAG-enhanced compilation architecture from \cite{pred} with three new layers: a persistent plant-context retrieval segment, a conversational interaction model, and a graduated tool-access mechanism.

\subsection{Conversational Engineering Interface}
The copilot presents a persistent, multi-turn chat interface. Unlike the single-query model of \cite{pred}, the engineer can engage in extended dialogue: asking the system to explain an existing program fragment, identify potential interlock conflicts, discuss design alternatives, or propose modifications before any code is generated.

\begin{figure}[htb!]
\centering
\includegraphics[width=1\linewidth]{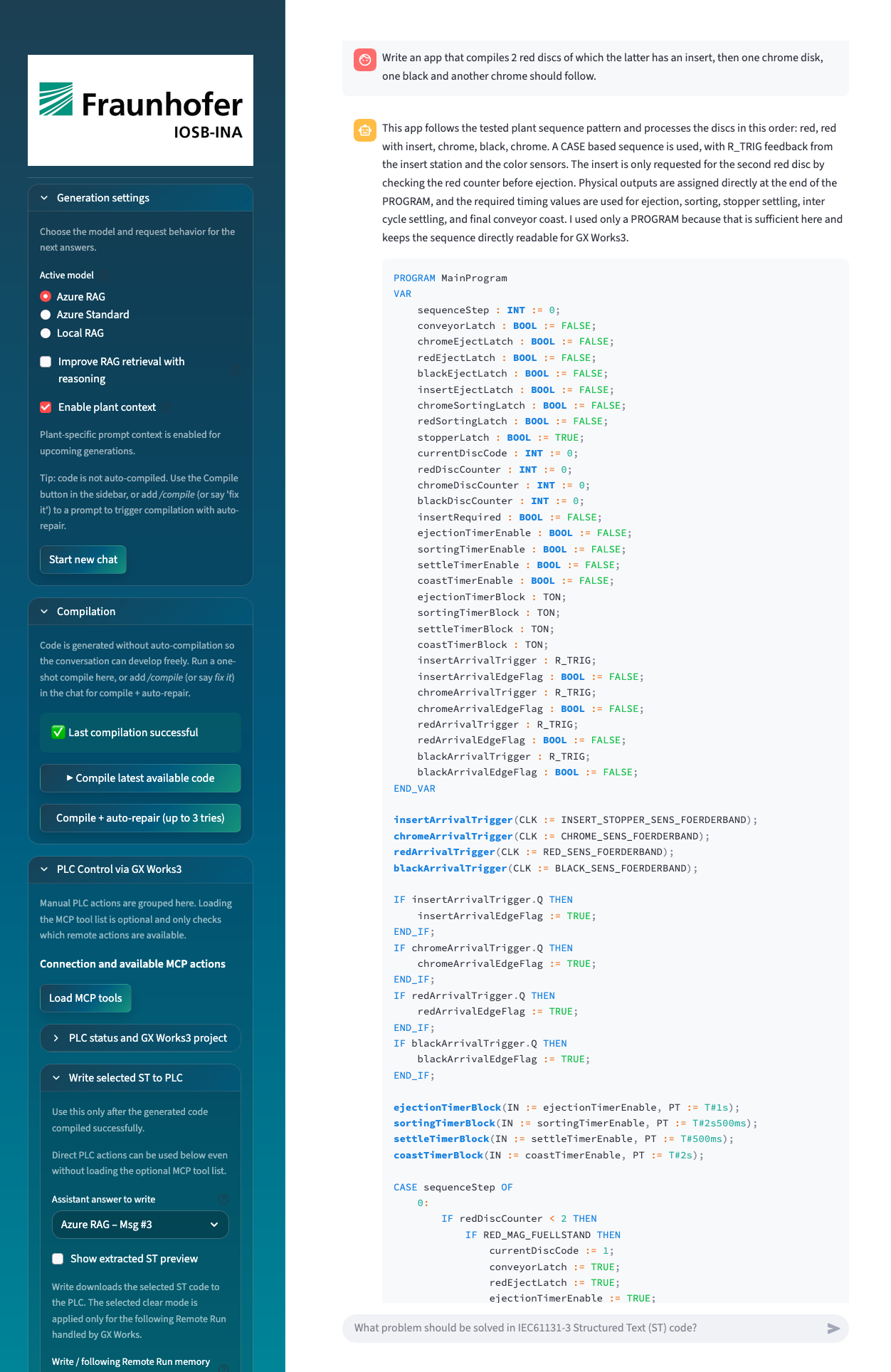}
\caption{The prototype web application used to control generation, compilation, and plant-aware context. The left panel exposes generation settings and MCP actions, while the main pane shows the conversation and generated ST code.}

\label{fig:webapp}
\end{figure}

Figure~\ref{fig:webapp} shows the web application that orchestrates the generation workflow, the plant context toggle, and the manual compile and deployment actions.

Code generation is explicitly request-triggered.
When the engineer asks for code, the model draws on all context layers; in subsequent turns the engineer can ask for modifications, and the model maintains the generated program as a shared artefact in the conversation.

\subsection{Plant Context Integration}
We define \emph{plant-awareness} as the system's ability to ground generated code in the target plant's I/O list,
label namespace, module configuration, and physical signal topology.
The key architectural addition over \cite{pred} is a dedicated plant-context retrieval segment.
On project initialization, the engineer provides:
\begin{itemize}
\item I/O list: signal names, addresses, data types, and descriptions for all inputs and outputs.
\item Device labels: the complete GX Works3 label table, including internal variables and inter-program signals.
\item Module configuration: rack layout, CPU type, and special-function module assignments.
\item Existing program fragments: current ST programs and function-block instances already deployed on the plant.
\end{itemize}

These materials are chunked, embedded, and indexed in a dedicated plant-context retriever,
separate from the general function-block and specification retrievers established in \cite{pred}.
At query time, a four-retriever strategy is used: library definitions, specification rules, general context, and plant-specific context.
Hence, similarity search operates within homogeneous chunk spaces and avoids cross-source interference.

\subsection{Graduated Tool Access}
Safety in industrial environments demands that LLM-generated artefacts are never silently applied to running machinery.
The copilot enforces four explicit levels of tool access:
\begin{itemize}
\item \textbf{Level 1 -- Advisory:} text only, no tool call.
\item \textbf{Level 2 -- Code generation:} code is displayed in the chat interface on explicit request.
\item \textbf{Level 3 -- Compilation:} the engineer triggers the compilation server and receives structured diagnostics.
\item \textbf{Level 4 -- Deployment:} the engineer explicitly confirms deployment to the PLC.
\end{itemize}

Each level transition produces real-world effects of increasing consequence.
The graduated structure ensures that advisory interactions carry no risk of unintended hardware modification.




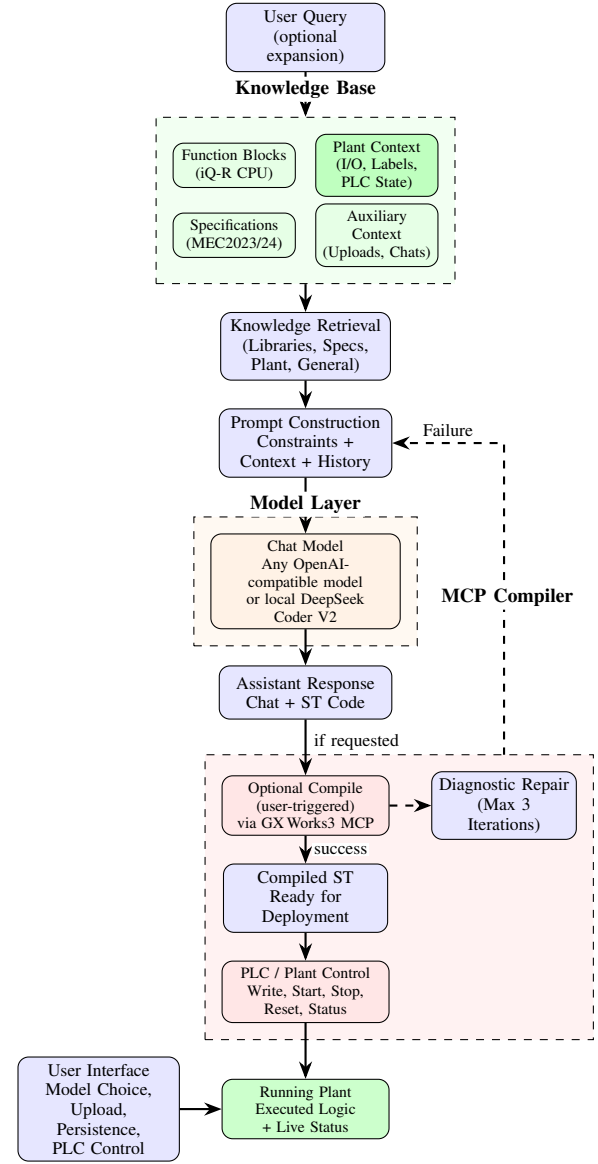
\begin{figure}[t]
\centering
\begin{tikzpicture}[scale=0.74, transform shape,
    block/.style={rectangle, draw, fill=blue!10, text width=2.85cm, text centered, rounded corners, minimum height=0.95cm, font=\small},
    datablock/.style={rectangle, draw, fill=green!10, text width=1.95cm, text centered, rounded corners, minimum height=0.72cm, font=\footnotesize},
    plantblock/.style={rectangle, draw, fill=green!25, text width=1.95cm, text centered, rounded corners, minimum height=0.72cm, font=\footnotesize},
    modelblock/.style={rectangle, draw, fill=orange!10, text width=3.15cm, text centered, rounded corners, minimum height=0.85cm, font=\footnotesize},
    compilerblock/.style={rectangle, draw, fill=red!10, text width=2.75cm, text centered, rounded corners, minimum height=0.8cm, font=\footnotesize},
    resultblock/.style={rectangle, draw, fill=green!20, text width=2.75cm, text centered, rounded corners, minimum height=0.85cm, font=\footnotesize},
    arrow/.style={-Stealth, thick},
    dashedarrow/.style={-Stealth, thick, dashed}
]

\node[datablock] (fb) at (0,0) {Function Blocks\\(iQ-R CPU)};
\node[plantblock] (plant) at (2.55,0) {Plant Context\\(I/O, Labels,\\PLC State)};
\node[datablock] (specs) at (0,-1.25) {Specifications\\(MEC2023/24)};
\node[datablock] (misc) at (2.55,-1.25) {Auxiliary Context\\(Uploads, Chats)};

\begin{scope}[on background layer]
\node[
    draw,
    dashed,
    fit=(fb)(plant)(specs)(misc),
    fill=green!5,
    inner sep=0.22cm
] (kb) {};
\end{scope}

\node[block, text width=2.6cm, above=0.80cm of kb] (user) {User Query\\(optional expansion)};

\node[block, text width=2.8cm, below=0.50cm of kb] (retrieval) {Knowledge Retrieval\\(Libraries, Specs, Plant, General)};

\node[block, below=0.50cm of retrieval] (prompt) {Prompt Construction\\Constraints + Context + History};

\node[modelblock, below=1.0cm of prompt] (model) {Chat Model\\Any OpenAI-compatible model\\or local DeepSeek Coder V2};

\node[block, below=0.65cm of model] (codegen) {Assistant Response\\Chat + ST Code};

\node[compilerblock, below=1.0cm of codegen] (compiler) {Optional Compile\\(user-triggered)\\via GX\,Works3 MCP};

\node[block, right=0.75cm of compiler, text width=2.35cm] (repair) {Diagnostic Repair\\(Max 3 Iterations)};

\node[block, below=0.50cm of compiler, text width=2.7cm] (compiled) {Compiled ST\\Ready for Deployment};

\node[compilerblock, below=0.50cm of compiled] (hardware) {PLC / Plant Control\\Write, Start, Stop,\\Reset, Status};

\node[resultblock, below=1.0cm of hardware] (output) {Running Plant\\Executed Logic + Live Status};

\node[block, left=0.75cm of output, text width=2.65cm] (ui) {User Interface\\Model Choice, Upload,\\Persistence, PLC Control};

\draw[arrow] (user) -- (kb.north);
\draw[arrow] (kb.south) -- (retrieval.north);

\draw[arrow] (retrieval) -- (prompt);
\draw[arrow] (prompt) -- (model);
\draw[arrow] (model) -- (codegen);
\draw[arrow] (codegen) -- node[pos=0.4, right=3pt, font=\small, fill=white, inner sep=1pt] {if requested} (compiler);
\draw[arrow] (compiler) -- node[midway, right=3pt, font=\small, fill=white, inner sep=1pt] {success} (compiled);
\draw[arrow] (compiled) -- (hardware);
\draw[arrow] (hardware) -- (output);
\draw[arrow] (ui) -- (output);

\draw[dashedarrow] (compiler) -- (repair);
\begin{scope}[on background layer]
\draw[dashedarrow] (repair) |- node[near end, above, font=\small] {Failure} (prompt);
\end{scope}

\begin{scope}[on background layer]
\node[
    draw,
    dashed,
    fit=(model),
    fill=orange!5,
    inner sep=0.22cm
] (cm) {};

\node[
    draw,
    dashed,
    fit=(compiler)(repair)(hardware),
    fill=red!5,
    inner sep=0.22cm
] (mcpbox) {};
\end{scope}

\node[fill=white, inner sep=2pt, rounded corners=0.25pt]
at ($(kb.north)+(0,14pt)$) {\textbf{Knowledge Base}};

\node[fill=white, inner sep=2pt, rounded corners=0.25pt]
at ($(cm.north)+(0,6.2pt)$) {\textbf{Model Layer}};

\node[fill=white, inner sep=2pt, rounded corners=0.25pt, align=center]
at ($(mcpbox.north)+(55pt,80pt)$) {\textbf{MCP Compiler}};

\end{tikzpicture}
\caption{System architecture for our ST coding assistant with plant-aware retrieval, user-triggered compilation and PLC control via MCP (based on \cite{pred}).}
\label{fig:architecture}
\end{figure}

\section{Demonstration and Preliminary Results}

\begin{figure}[t]
\centering
\includegraphics[width=\columnwidth]{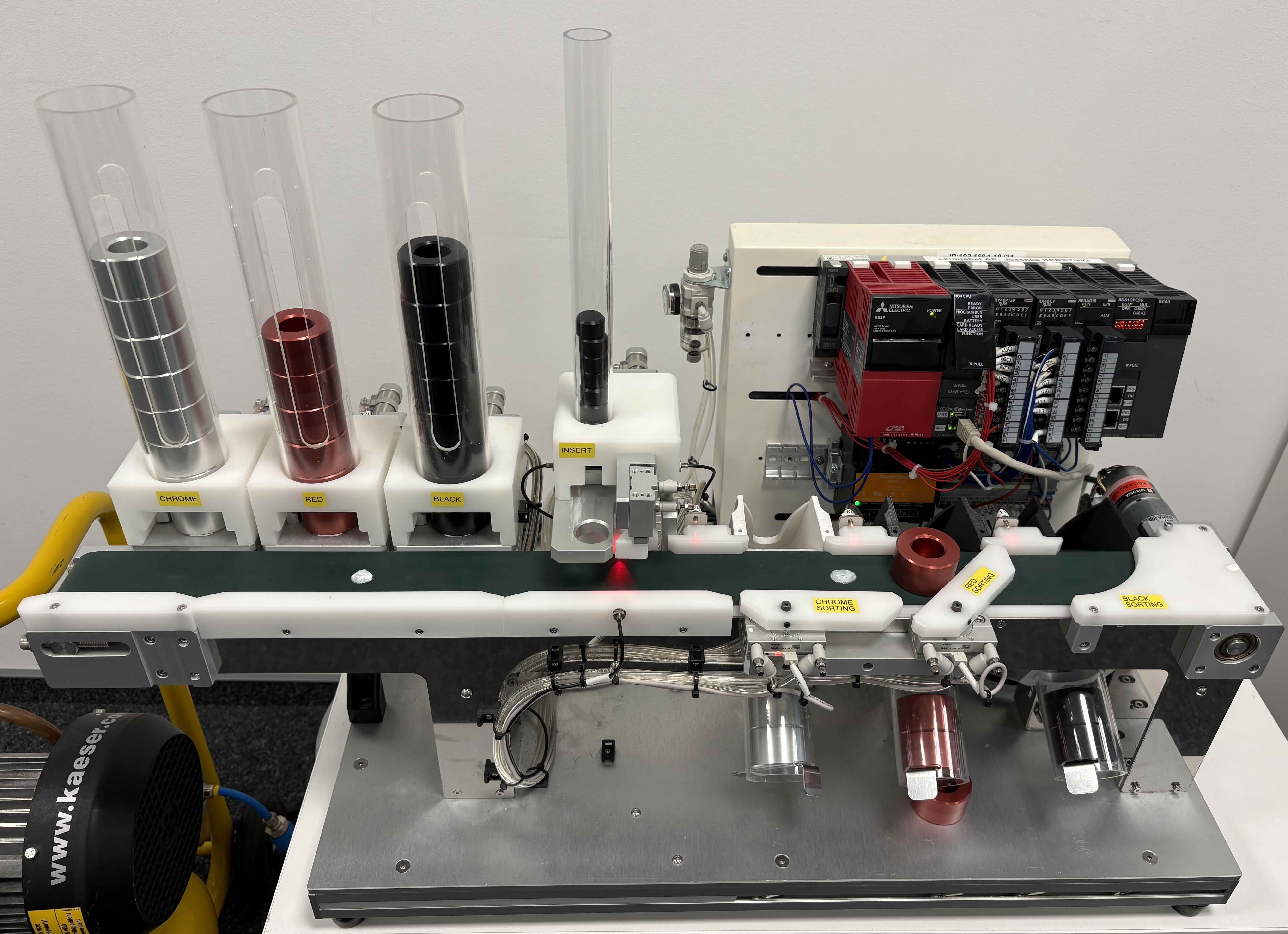}
\caption{Demonstration plant setup at a research test facility providing metal discs in different colors, inserts, sensors, and actuators for sorting them. The plant is equipped with a Mitsubishi Electric iQ-R series CPU (model R04).}

\label{fig:demo-plant}
\end{figure}

\subsection{Setup}
Figure~\ref{fig:demo-plant} shows the physical setup used in the demonstration.
The demonstration plant is a sorting machine that, based on air pressure and light sensors, can push metal discs of different colors onto a conveyor belt. The colors are chrome, red and black. Since these colors reflect differing amounts of light, the sensors can be configured to detect them with different thresholds.
Another actuator with sensors can put inserts into the discs. Users may configure their own sorting app that configures a mixture of colored discs with or without inserts using AI. The plant can thus be reconfigured based on users' current needs. The prototype may serve as a demonstration for more complex plants, e.g., in the automotive industry.

The demo setup uses a conveyor belt that transports discs to the right where, having passed all magazines, they are sorted into three boxes.
Chrome, as the most reflective color, comes first, followed by red and black.
Chrome and red are actively sorted by a pneumatic cylinder that pushes them off the belt, while black discs are passively sorted by gravity.
In Fig.~\ref{fig:demo-plant}, the plant is pushing red off the conveyor belt.

The plant is equipped with a Mitsubishi Electric iQ-R series CPU programmed via a prototype web interface and GX Works3. Plant context was ingested as an I/O list, a label table, and existing ST program fragments. Moreover, the actual plant setup had to be explained to the model, since the I/O list and labels alone do not capture the physical layout and signal semantics. That is, the model had to be informed where magazines are, when they are loaded (i.\,e. the corresponding sensor must be triggered as TRUE) and that the discs are pushed on a conveyor belt to the right, where actuators in a certain order are situated.

\subsection{Example Interaction Trace}
The following excerpt shows the intended interaction pattern for plant-aware code generation and validation.

\begin{quote}
\textbf{Engineer:} Are there blue discs available?

\textbf{Copilot (Advisory -- L1):} No. The plant context defines only chrome, red, and black discs. There is no blue disc magazine, sensor, or sorting path.

\textbf{Engineer:} Generate the sequence red, red with insert, chrome, black with insert, chrome.

\textbf{Copilot (Code gen -- L2):} The generated Structured Text uses the plant's actual labels and control flow, including the black-disc branch gated by \texttt{BLACK\_MAG\_FUELLSTAND} and \texttt{INSERT\_MAG\_FUELLSTAND}.

\textbf{Engineer:} [Triggers compilation]

\textbf{Copilot (Compile -- L3):} GX Works3 compilation successful.

\textbf{Engineer:} Can the black disc also receive an insert?

\textbf{Copilot (Code gen -- L2):} Yes. The same insert-station logic applies to the black disc before it continues on the black path. Code: ...

\textbf{Engineer:} [Triggers compilation]

\textbf{Copilot (Compile -- L3):} GX Works3 compilation successful.

\textbf{Engineer:} [Triggers deployment to PLC, chooses to delete latches and variable status; Starts PLC run]

\textbf{Copilot (Deploy -- L4):} Code deployed to PLC. The plant is now executing the new sequence. \textit{See Fig.\,\ref{fig:demo-plant}}.
\end{quote}

\section{Discussion \& Conclusion}

This paper presented a plant-aware conversational copilot for PLC engineering that extends the RAG-enhanced code generation architecture of prior work \cite{pred} with persistent plant-context ingestion, a chat-first multi-turn interaction model, and a graduated tool-access ladder from chat to deployment. By retrieving plant-specific signal names at query time, the copilot is intended to generate code that is more appropriate for the target plant's namespace. A structured quantitative evaluation comparing plant-aware and generic-context generation on matched task sets is planned.

\textbf{Limitations} include that the current system is demonstrated on a single plant within Mitsubishi Electric's iQ-R ecosystem. Cross-vendor generalizability is not in the scope. An evaluation methodology should capture functional correctness, which requires hardware-in-the-loop testing or domain-expert review beyond compilation success. Human evaluation from a single organizational context may limit generalizability.

\textbf{Future Work} includes (A) execution feedback connecting PLC runtime signals and the copilot context,
(B) multi-PLC orchestration for cross-controller installations,
(C) graphical IEC 61131-3 extensions for Ladder diagram generation and legacy AWL code, and (D) a quantitative evaluation across multiple industrial-size plants with plant-awareness metrics.


\bibliographystyle{IEEEtran}
\bibliography{refs}

\end{document}